# Decision Making for Autonomous Vehicles

Xinchen Li, Levent Guvenc, Bilin Aksun-Guvenc

Automated Driving Lab, Department of Mechanical and Aerospace Engineering, Ohio State University

**Abstract**

This paper is on decision making of autonomous vehicles for handling roundabouts. The round intersection is introduced first followed by the Markov Decision Processes (MDP), the Partially Observable Markov Decision Processes (POMDP) and the Object Oriented Partially Observable Markov Decision Process (OOPOMDP). The Partially Observable Monte-Carlo Planning algorihtm (POMCP) algorithm is introduced and OOPOMDP is applied to decision making for autonomous vehicles in round intersections. Decision making is formulated as a POMDP problem and the penalty function is formulated and set followed by improvement of decision making with policy prediction. The augmented objective state and policy based state transition is introduced simulations are used to demonstrate the effectiveness of the proposed method.

## 1. Introduction

Self driving or autonomous vehicles are already available in limited scale and are expected to become more available in the near future. When autonomous vehicles also use on-board units which are vehicular to everything communication modems, they become connected and autonomous vehicles (CAV) [1]. Due to their increasing availability, CAVs have been the focus of both academic and industry research for a while and, as a result, there are is lot of research on autonomous driving function controls [2], [3], [4], [5], [6], [7], [8], [9], ;10], [11] and their higher level decision-making algorithms [12], [13], [14], [15]. Planning and decision making are the core functions for an autonomous vehicle for driving on the road safely and efficiently under different traffic scenarios. As discussed in [16], the decision making and planning algorithms for autonomous vehicles are aiming at solving significant problems in autonomous driving, like (a) determining the future path, (b) utilizing observations of the surrounding environment from the perception system, (c) acting properly when interacting with other road users, (d) instructing low-level controller of the vehicle and (e) ensuring autonomous driving is safe and efficient. Therefore, the planning and decision making affects the autonomous vehicle decisively. Depending on the traffic scenario, autonomous driving functions are designed for highway driving, off-road driving and urban driving. The research for highway driving and off-road driving have been going on for a long time and with many results on planning and decision making. Yet, due to the complexity of the urban traffic scenario, the decision making and planning for urban traffic environment has always been very challenging with many unsolved problems.

The complexity of the urban traffic scenario is manifested in the following aspects which are discussed next. (a) The various of road types. Roads on a highway scenario are similar, vehicles should stay in lane or execute lane change based on the drivers demand. Unlike highway, an urban traffic scenario is composed of different road types, including lanes, intersections, traffic circles, roundabouts as well as lanes for bicyclists and cross-walks for pedestrians. Hence, the decision making and planning cannot only take care of vehicle behavior such as lane following and lane changing, the merging into traffic circle/intersection or encountering lanes for other road users also need to be considered. (b) Different types of road users share the road. In an urban traffic scenario, the road users are not only vehicles, but also other vulnerable road users (VRU) such as bicyclist and pedestrians. It will cause much more severe damage when VRUs are involved in traffic accidents. The safety of driving, thus, is the one thing that autonomous vehicles and the planning algorithm always prioritizes. (c) Traffic signals are involved but absent sometimes. There are not always traffic signals in urban traffic scenarios. The autonomous vehicles needs to adjust to the occasion where no traffic signals are involved for traffic control. Those points make urban traffic complex and difficult to deal with. One of the major challenges for autonomous vehicles on decision making in urban traffic scenario is the part for intersections and roundabout intersection, especially if they are not signalized.

One of the focuses in this paper is the decision making and planning for round intersections. A round intersection is a special case of unsignalized intersections. An intersection is an junction where roads meet and cross. Intersections, based on the existence of traffic signals, are usually categorized into two types: signalized intersections and unsignalized intersections. The signalized intersection is a centralized control system. The traffic flow at signalized intersection is controlled by the traffic signal, either traffic lights or traffic signs, thus the vehicles will behave according to the traffic signal, not requiring extra decision making or behavior planning. In contrast, at an unsignalized intersection, the decision making and planning is decided upon the driver or the planner for an autonomous vehicle for determining the behavior when approaching the intersection as well as interacting with other road users nearby or the traffics within the intersection zone.

The organization of the rest of the paper is as follows. The round intersection is introduced in Section 2. The Markov Decision Processes (MDP) and the Partially Observable Markov Decision Processes (POMDP) are introduced in Section 3. The Object Oriented Partially

Observable Markov Decision Process (OOPOMDP) is introduced in Section 4. The Partially Observable Monte-Carlo Planning algorihtm (POMCP) algorithm is given in Section 5. OOPOMDP is applied to decision making for autonomous vehicles in round intersections in Section 6. In Section 7, decision making is formulated as a POMDP problem. The penalty function is formulated and set in Section 8. Improvement of decision making with policy prediction is presented in Section 9. The augmented objective state and policy based state transition is introduced in Section 10. Simulations are used in Section 11 to demonstrate the effectiveness of the proposed method. The paper ends with conclusions and recommendations.

## 2. Round Intersections

A round intersection is a special case of unsignalized intersection. A typical geometry layout of a round intersection is as shown in Figure 1. A round intersection can be viewed as combinations of three types of roads, the one lane straight roads before the round intersection, the entering/exit part of the round intersection and the round intersection in which the vehicle can only drive in the counter-clockwise direction as shown in Figure 1.

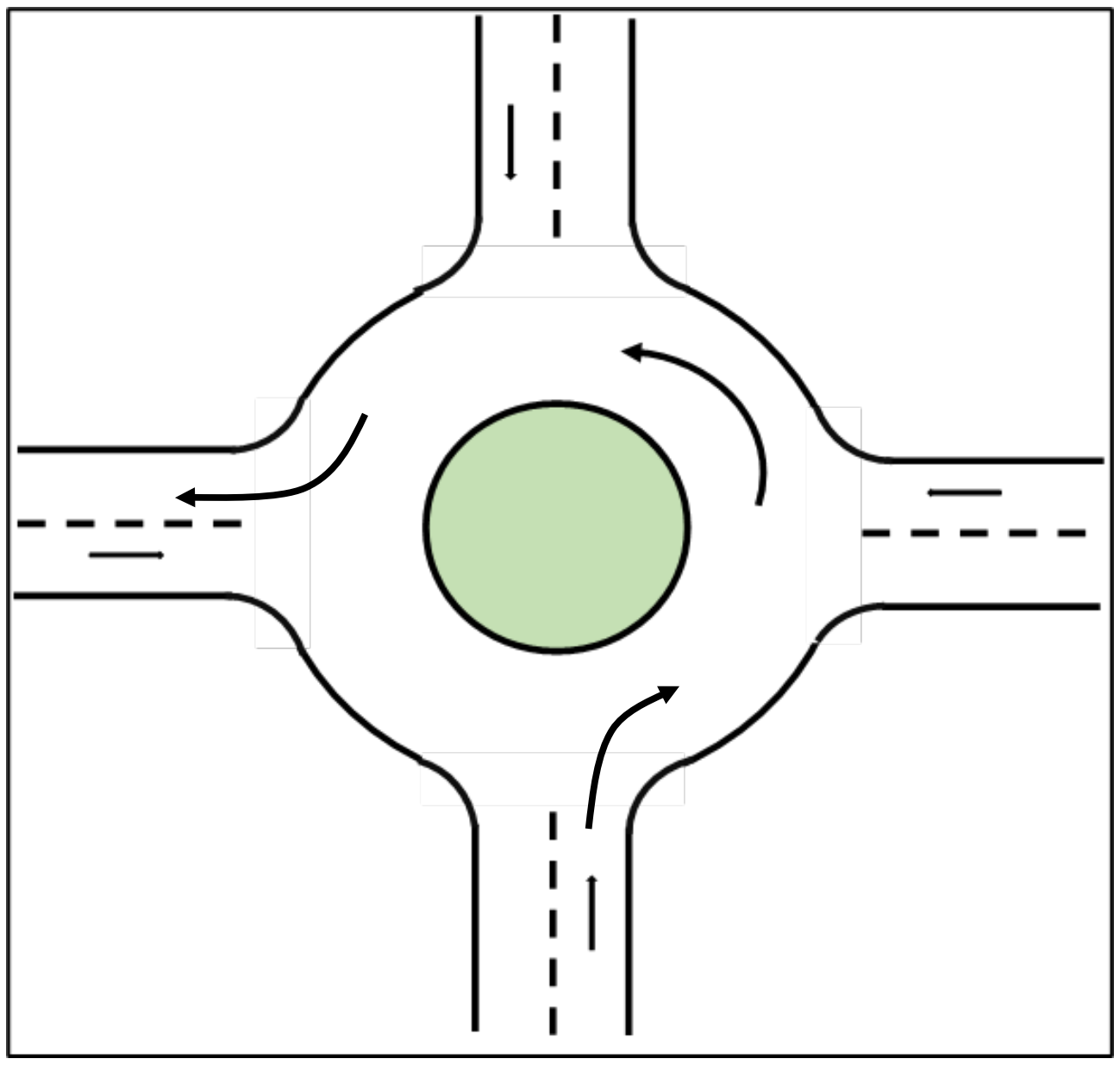

Figure 1 Diagram of a Round Intersection

Vehicles enter the round intersection from the straight road or exit it in the area of exiting or entering, and vehicles will follow the lane in straight roads and in the round intersection. For a single vehicle, the change of angular velocity in different parts of the road make it difficult for the vehicle motion model to track the vehicle motion behavior. When there are multiple vehicles in the traffic scenario, the interaction between vehicles requires a good decision making and planning algorithm for the autonomous vehicle.

In this paper, a decision making algorithm based on Partially Observable Markov Decision Making Process (POMDP) for planning acceleration of the autonomous vehicle is proposed. With the involvement of policy prediction method, the tracking of vehicle trajectories of other vehicles for the planning horizon in different area of the traffic scenario are dealt for better decision making.

## 3. POMDP for Decision Making and Planning

Motion planning and decision making problem of autonomous vehicles can be viewed as a sequential decision making problem. Markov Decision Processes (MDP) and the Partially Observable Markov Decision Processes (POMDP) represent a very wide range of methods for solving the sequential decision making problem. In the Markov Decision Process, an agent takes actions that affect the whole system which includes the environment and the agent itself. The agent is looking for actions which lead to expected maximal rewards collected from the whole system as illustrated in Figure 2.

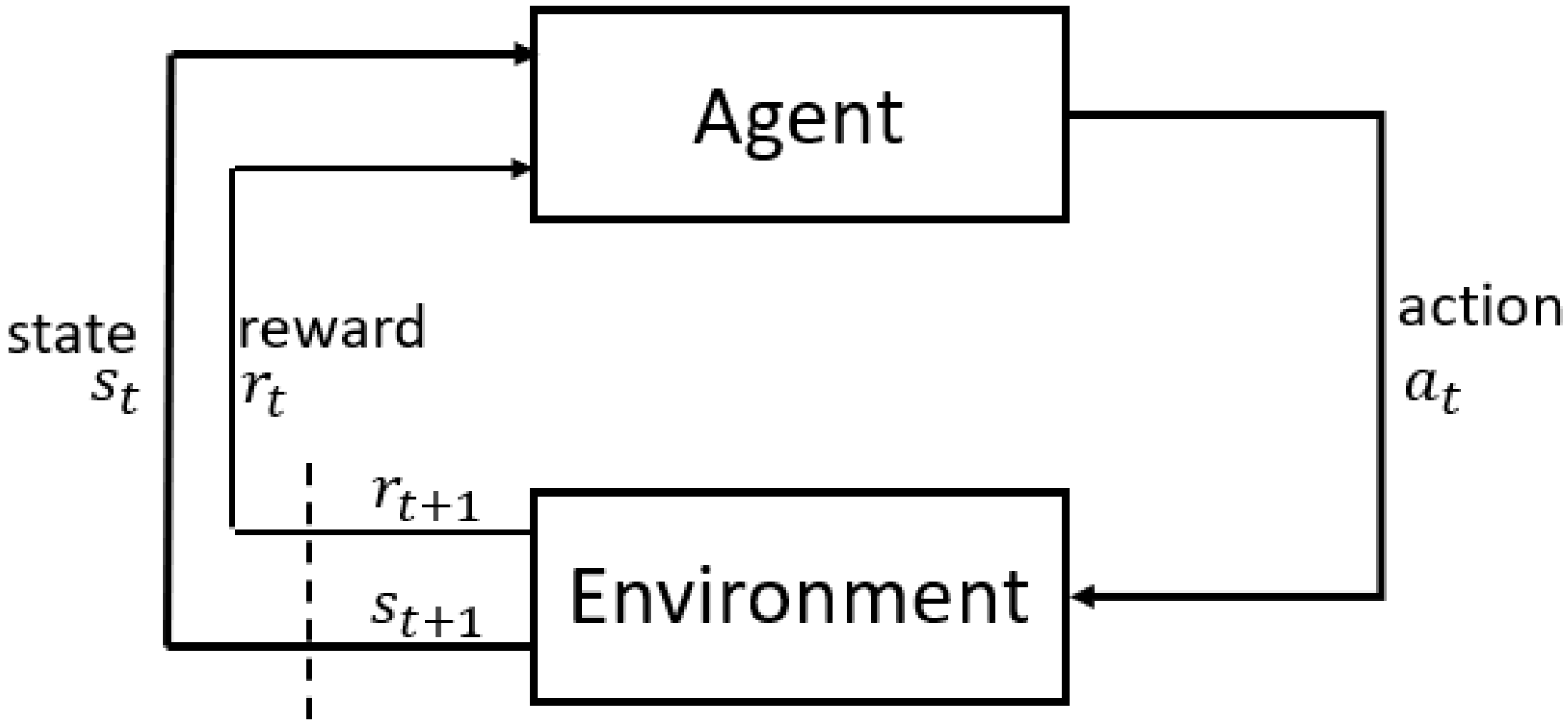

Figure 2 Diagram of Sequential Decision Making Problem

Formally, the MDP is defined by a 5-tuple $\langle S, A, T, R, \gamma \rangle$, where $S$ represents the state space in which there are states $s$ representing possible states of the world, including the states of the agent as well as uncontrollable factors in the environment. $A$ represents the action space, which is a set of executable actions that the agent can take. $T$ is the transition model, $R$ is the reward function with $\gamma$ to be the discounted factor. The transition model $T(s'|s,a)$ denotes the probability that the system updates to state $s'$ given that the system takes an action $a$ at the state $s$. The reward function $R(s,a)$ provide the immediate reward when system takes an action $a$ at the state $s$. The action $a$ is derived from the policy function $\pi(s,a) = P(a|s)$ to achieve the overall optimization goal of the cumulative expected reward:

$$E[\sum_{t=0}^{\infty}\gamma^t R(s_t,a_t)] \tag{1}$$

For the problem of POMDP, the difference of it from the problem of MDP is that the system state $s$ is not fully observable to the agent. Instead, the agent can only access observations that are generated probabilistically based on the action and the possible true states. With this issue, the PODMP can be represented as a 7-tuple extended from MDP, $\langle S, A, T, R, O, Z, \gamma \rangle$, where $S, A, T, R$ and $\gamma$ have the same meanings as in an MDP. The two additional elements, $O$ represent the observation space which contains observations $o$ that the agent perceives from the system and the environment and $Z$ is the observation model, as $Z(o|s',a,s)$ denotes the probability or probability density of receiving observation $o$ in state $s'$ given the previous state and action as $s,a$.

The distinction between POMDP and MDP is that the agent does not have a full knowledge of the environment and the system. Hence, the state information has to be inferred from the entire history of previous actions and observations as well as the initial belief information $b_0$. The policy is now a mapping from history to an action. The history at time t is denoted as $h_t = \{b_0, a_0, o_1, a_1, o_2, ..., a_{t-1}, o_t\}$. As a contrast, the policy mapping function now depends on the history rather than the states themselves, $\pi(h_t, a) = P(a \mid h_t)$. Due to the lack of knowledge of state $s$, belief state is introduced that is the probability distribution over states given the history.

$$b(s) = \Pr(s_t \mid h_t = h) \tag{2}$$

Which is important for the POMDP for updating the belief states. After reaching a new state $s'$ the agent observes a new observation $o$ based on the observation model. The belief update rule for POMDP is,

$$b'(s') = \eta Z(o \mid s', a) \sum_{s \in S} T(s' \mid s, a) b(s) \tag{3}$$

The overall optimization is refined as

$$a^* = \arg\max_{a_t} E[\sum_{t=0}^{\infty} \gamma^t R(s_t, a_t) \mid b_0] \tag{4}$$

In the case of autonomous driving decision making problem, the ego-vehicle does not have full observation over other vehicle or road user states. Hence, the POMDP provides a better method for solving the sequential decision making problem for autonomous driving under uncertainties and partial observability.

## 4. Object-Oriented POMDP

Traditional POMDP works well for problems with state space which is of small domain and low dimension. However, the POMDP becomes computationally intractable for planning over large domain and brings up the “curse of dimensionality” issue for its solution [17]. Since the state space is not fully observable for the agent, the agent needs to develop a belief representation over the state space which grows exponentially with the number of the state variables. As for the decision making problem of autonomous vehicle driving, the number of other vehicles in the environment will normally be large, and the decision horizon can be long to finish a driving task in specific traffic scenarios. Both of factors in a traffic scenario lead to the large domain for the POMDP problem.

To improve the performance and deal with the intractability of POMDP over a large domain, an Object-Oriented POMDP is proposed in [18] for solving a multi-object search task. It is also useful when applied to solving autonomous driving decision making problem due to the large domains in the problem formulation of a traffic scenario with multiple vehicles in it.

Object-Oriented POMDP (OOPOMDP) provides an extension for POMDPs by factorizing both state and observation spaces in terms of object. It uses object states abstractions to enable the agent to reason about objects and relations between objects. The OOPOMDP problem can be represented as a 11-tuple $\langle C, Att(c), Dom(a), Obj, S, A, T, R, Z, O, \gamma \rangle$. In this notation, the $S, A, T, R, O, Z$ and $\gamma$ have the same meanings as those in the POMDP problems. However, in the problem of OOPOMDP, a new set $Obj$ is introduced. $Obj = \{obj_1, ..., obj_n\}$ is a set that the state space $S$ and observation space $O$ are factored into. Each $Obj_i$ is an instance of particular class, $c \in C$. Those classes have their class specific attribute that are defined in the set $Att(c)$. $Dom(a)$ defines the possible values of each attribute in the attribute set. The dual factorization of S and $O$ allows the observation functions to exploit shared structure so as to define observations grounded in the state with varying degrees of specificity: over a class of objects, a single object, or an object's attribute.

One of the important steps in solving a POMDP that causes the problem of "curse of dimensionality" is the belief update procedure, as shown in Equation (3). The belief state over possible state space has to be updated according to the transition and observation models. The state space grows exponentially along the update of the belief space regarding the number of objects.

Assume that there are $M$ possible states for the $N$ objects in the current decision region. Then states to be covered by the POMDP are:

$$|S| = \prod_{i=1}^{N} |S_{obj_i}| = |M|^N \tag{5}$$

Therefore, the dimension of the belief state $b(s)$ also grows exponentially with the state space.

The OOPOMDP method improves the step for belief update over multiple objects and reduces the dimension of states by exploring one possible independence assumption. When assuming that objects in the environment is independent from each other, the belief state over the state space can be viewed as a union of belief states over the state of each object, as shown in Equation (6),

$$\begin{aligned} b(s) = b(\bigcup_{i=1}^{N} s_i) &= b(s_1)b(s_2)...b(s_N) \\ &= \prod_{i=1}^{N} b(s_i) \end{aligned} \tag{6}$$

Thus, the exponentially growing belief state now scales linearly in the number of objects $N$ in the environment. The dimension of the unfactored belief, which is the belief that are not processed by the OOPOMDP is $|M|^N$. While the dimension of the factored belief state is $N|M|$. When providing object-specific observation $o_i$ with respect to each of the independent object in the environment, the belief update can also be object-specific for object *i,* as shown in Equation (7).

$$b_i'(s_i) = \eta p(o_i | s_i) b_i(s_i) \tag{7}$$

where *i* for the $i^{th}$ object and $\eta$ is the normalization factor for the belief state within the range of $[0,1]$.

## 5. The Object-Oriented POMCP Algorithm

To solve the OOPOMDP problem, a modified version of famous Partially Observable Monte-Carlo Planning algorihtm (POMCP) algorithm [19] is proposed. The POMCP algorithm is a well-known online POMDP planning algorithm that shows significiant success over large domain POMDP problems. POMCP applies Monte-Carlo Tree search to find the optimal policy and best action by estimating the Q-value over a certain action and next belief state and is shown in Table 1.

| Algorithm: Partially Observable Monte-Carlo Planning (POMCP) | |
|---|---|
| **Procedure** Search (h):<br>Repeat:<br>if $h = empty$ :<br>$s \sim I$<br>else:<br>$s \sim B$<br>end if<br>SIMULATE (*s,h,0*)<br>Until Timeout()<br>return $\arg\max_b V(hb)$<br>**end Procedure**<br><br>**Procedure** ROLLOUT (*s, h, depth*):<br>if $\gamma^{depth} < \epsilon$ :<br>return 0<br>end if<br>$a \sim \pi_{rollout}(h, \cdot)$<br>$(s', o, r) \sim g(s, a)$<br>return $r + \gamma \text{ROLLOUT}(s', hao, depth+1)$<br>**end Procedure** | **Procedure** SIMULATE(*s, h, depth*):<br>if $\gamma^{depth} < \epsilon$ :<br>return 0<br>if $h \notin T$ :<br>for $\forall a \in A$ :<br>$T(ha) \leftarrow (N_{init}(ha), V_{init}(ha), \phi)$<br>end for<br>return ROLLOUT(*s, h, depth*)<br>end if<br>$a \leftarrow \arg\max_b V(hb) + c\sqrt{\frac{\ln N(h)}{N(hb)}}$<br>$(s', o, r) \sim g(s, a)$<br>$R \leftarrow r + \gamma \text{SIMULATE}(s', hao, depth+1)$<br>$B(h) \leftarrow B(h) \cup \{s\}$<br>$N(h) = N(h) + 1$<br>$N(ha) = N(ha) + 1$<br>$V(ha) = V(ha) + \frac{R - V(ha)}{N(ha)}$<br>return $R$<br>**end Procedure** |

Table 1 POMCP Algorithm (source: [19])

A search tree diagram starting from current history $h$ with two actions $a_1$ and $a_2$ is presented in Figure 3. Each search tree starts from current history $h$, and forward simulates with two actions. New observations will be generated based on the transition and observation models so that new tree nodes can be added to the tree.

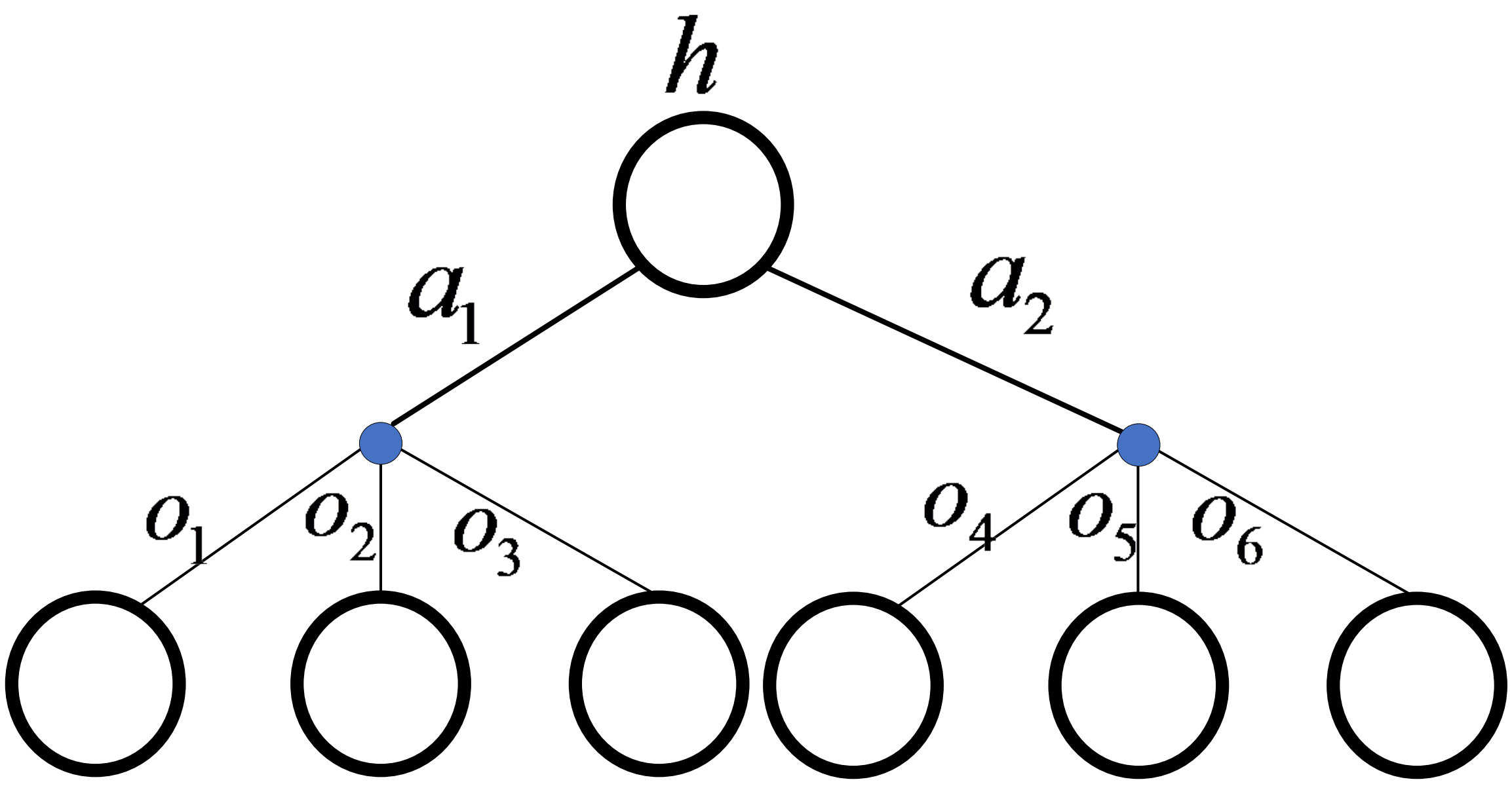

Figure 3 Diagram of Search Tree of Two Actions

In a search tree of POMCP, each tree node represent a tuple of $\langle N(h), V(h), B(h) \rangle$ with respect to the current history $h$. $N(h)$ represents the number of the current node being visited. $V(h)$ represents the node value corresponding to the history of the node. When new action and observations are generated, the history is also updated, noted as concatenated to the current history. $ha$ means new action is added to the history and $hao$ means the new action and observation are both added to the history $h$, and the cumulative reward averaged by the number of visits, given by

$$V(h)=\frac{\sum R_i}{N(h)}=\frac{\sum_{k=t}^{\infty}\gamma^k r_k}{N(h)} \tag{8}$$

The value is related to the cost function setting in the problem which the POMCP algorithm is trying to solve. The $B(h)$ of the node, the set of particles of the states, is the feature that differentiate POMCP from normal Partially Observable Upper Confidence Tree (PO-UCT). In POMCP, the belief state is approximated by a set of particles sampled from the history. After the sampled particles reach the expected number, a Monte-Carlo procedure to update the particles in the set is taken for updating the particles in the set as well as sampling corresponding observations, rewards and state transitions. Each of the tree search cycle begins with sampled initial state from the particle sets. The Objective-Oriented POMCP algorithm [20] is introduced in Table 2.

For each decision cycle, a forward search tree $T$ is constructed by iteratively sampling state particles from the current belief states over the object states for the procedure SIMULATE. Each of the tree node in the search tree contains a tuple of information related to corresponding history $\langle N(h), V(h)\rangle$. In the SIMULATE procedure, the forward simulation to grow the tree is done by the simulator $G(s,a)$ and it works with the sampled state particles to generate simulated observation $o$, the direct reward $r$ and the next state $s'$. This could be an explicit model or a black-box simulator.

| Algorithm: Objective-Oriented Partially Observable Monte-Carlo Planning | |
|---|---|
| **Procedure** Search $(h)$ :<br>$b \leftarrow$ Prior<br>$s \leftarrow$ InitialState<br>Repeat:<br>$\hat{s} \sim$ SAMPLE$(b)$<br>SIMULATE$(\hat{s},\{\},0)$ | **Procedure** SIMULATE (*s, h, depth*)<br>if $\gamma^{depth} < \epsilon$ :<br>return 0<br>if $h \notin T$ :<br>for all $a \in A$ |
| Until Timeout()<br>$a \leftarrow \arg\max_{a} V(ha)$<br>$s', o, r \leftarrow \varepsilon(s,a)$<br>$b' \leftarrow$ UPDATE$(b,a,o)$<br>**end Procedure**<br><br>**Procedure** UPDATE(*b, a, o*):<br>for $obj_i \in obj$:<br>$b_o'(s') \leftarrow \eta Z(o'\mid s',a)\sum_{s\in S} T(s'\mid a,s) b_o(s)$<br>return $\prod b_o'(s')$<br>**end Procedure**<br><br>**Procedure** SAMPLE(*b*):<br>for $obj_i \in obj$:<br>$\hat{s}_o \sim b_o$<br>return $\bigcup \hat{s}_o$<br>**end Procedure** | $T(ha) \leftarrow (N_{init}(ha), V_{init}(ha), \phi)$ e<br>end for<br>return Rollout($s, h, deph$)<br>end if<br>$a \leftarrow \arg\max_b V(hb) + c\sqrt{\frac{\ln N(h)}{N(hb)}}$<br>$(s',o,r) \sim g(s,a)$<br>$R \leftarrow r + \gamma$SIMULATE$(s', hao, depth+1)$<br>$N(h) = N(h)+1$<br>$N(ha) = N(ha)+1$<br>$V(ha) = V(ha) + \frac{R - V(ha)}{N(ha)}$<br>**end Procedure** |

Table 2 OOPOMCP Algorithm (source: [20])

If a history sequence $h$ is first encountered during the tree search, then a new node corresponding with $h$ is added to the tree with a new node initialized and the ROLLOUT procedure will be performed for the leaf nodes in the tree. Otherwise, a discounted reward will be added up together and will keep the simulation going on. During tree search, the selected action branch is based on the UCB1 policy,

$$a \leftarrow \arg\max_{a} V(ha) + c\sqrt{\frac{\log N(h)}{N(ha)}} \tag{9}$$

Instead of using a set of particles to represent the belief state and update the particle sets during SIMULATE, the belief state update for OOPOMCP is specific and explicit for each object. After an optimal action is selected for maximizing the UCB1 value from the tree search, the real action will be taken by the agent and obtain real observations from the environment. The real action as well as the real observation updates the belief and end the decision cycle. According to [20], the explicit belief update can also help with dealing with the degeneration of particles in POMCP. Due to the large probability of rejecting particles, sampling enough particles into the set for approximating belief states becomes difficult. Degeneration of particles takes places when successively computing the belief states. It is harmful when the belief state $b$ does not contain true environment states. Resampling cannot recover either when the belief state particles fails to represent the true environment states.

The belief update according to real action and real observation guarantees the correctness of belief state $b$ representing the real world and will not cause the degeneration of the particles. This allows the OO-POMCP algorithm to enable robust planning in large domains with multiple objects.

## 6. Decision Making for Autonomous Vehicles for Round Intersection using the OOPOMDP

The goal of implementing the decision making algorithm at a round intersection is to realize a safe, comfortable and efficient driving of an autonomous vehicles that interacts with multiple other vehicles. To achieve this goal, not only is a decision making algorithm necessary but also a good path planning is also required. However, it is out of the scope of this section and we will not discuss the path planning method here. The assumption of this problem is that the vehicle will drive on a designated path through a round intersection.

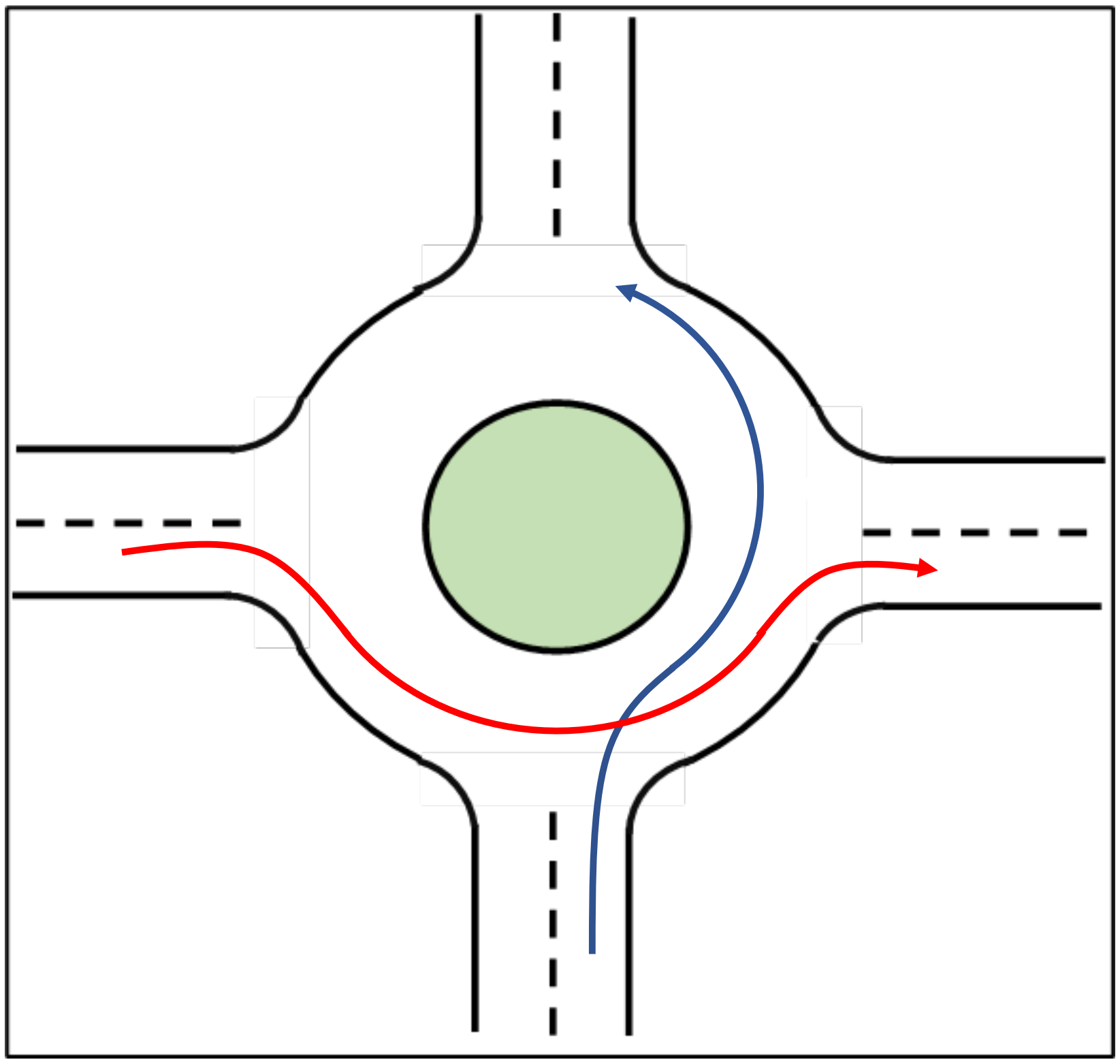

Figure 4 Diagram of Two Vehicles Passing through the Round Intersection

As shown in Figure 4, in the current situation, two vehicles are passing through the round intersection. The blue line represent the path for the ego autonomous vehicle and the red line shows the path for another vehicle which could be a human driver vehicle or autonomous vehicle. These two paths has an intersection. Thus, to make sure the collision does not happen at the intersection point between two paths and the ego-vehicle can pass through the round intersection within a shorter time as compared with executing stop-and-pass policy, the decision making algorithm for a round intersection is necessary.

The vehicle model is an important factor for the vehicle control in autonomous driving. Even though there is no perfect model to describe how the vehicle is moving, there are several vehicle models commonly used since they show good approximations and performance for vehicle control problem. An example is the single-track model, and the dynamic model for analyzing vehicle dynamics.

The autonomous vehicle control is a typical task that can be modeled in a hierarchical architecture. High level control as for decision making and low level control for path following as well as the control for brake and throttle together make up the vehicle control so that autonomous vehicle can act due to the control command and the planned acceleration or steering. In this work, as we are mainly concentrating on high level decision making part for the vehicle control and a model which describes the vehicle kinematic motion well is enough. Hence, we are using the kinematic motion model in discrete time domain at a round intersection given the assumption that the vehicle can be viewed as a block in 2-D plane. The vehicle model for the ego vehicle with the subscript of $r$ is denoted as

$$S_{t+1} = T_r(S_t, a_t) \tag{10}$$

where $S_t$ represents the state vector of the vehicle and $a_t$ is the input to drive the model.

Model (10) is expanded in detail as,

$$x(t+1) = x(t) + v(t)\cos\theta(t)\Delta t \tag{11a}$$

$$y(t+1) = y(t) + v(t)\sin\theta(t)\Delta t \tag{11b}$$

$$v(t+1) = v(t) + a(t)\Delta t \tag{11c}$$

$$\theta(t+1) = \theta(t) + w(t)\Delta t \tag{11d}$$

In Equations 11, $x(t), y(t)$ describes the 2D planar position, $\theta(t)$ is the heading angle of the vehicle model with respect to a fixed coordinate system. $v(t)$ is the linear velocity of the vehicle. The input $a_t$ that drives the vehicle is $\langle a(t), w(t) \rangle$, the linear acceleration and the angular velocity or the yaw rate of the vehicle. The acceleration determines how fast the vehicle can change speed and its direction of the vehicle. $\Delta t$ is the time-step for the discrete model. In this work, the above model is used for updating the states of the agent. Since the agent has no information of other vehicles, we assume that other vehicles will follow the similar model yet no speed change or direction will happen, denoted as

$$S_{t+1} = T_v(S_t) \tag{12}$$

can be expanded as,

$$x(t+1) = x(t) + v(t)\cos\theta(t)\Delta t \tag{13a}$$

$$y(t+1) = y(t) + v(t)\sin\theta(t)\Delta t \tag{13b}$$

$$v(t+1) = v(t) \tag{13c}$$

$$\theta(t+1) = \theta(t) \tag{13d}$$

The positioning information, the heading angle and the velocity of other vehicles will be collected at each time a perception snapshot is made.

In this paper, the input of the ego vehicle is selected from a finite set of actions, $\{a_1, a_2, ..., a_m\} \times \{w_1, w_2, ..., w_k\}$ for linear acceleration and angular velocity. At every decision step, the ego vehicle will take a linear acceleration for determining its velocity and an angular velocity for determining its direction of driving.

The change of direction of the yaw rate introduces the discontinuity of tracking the vehicle motion using the model in 10 and will influence the decision making for the forward Monte-Carlo simulation part and will be discussed in detail later.

## 7 Formulation of Decision Making as a POMDP Problem

In this paper, without loss of generality, the road users are all considered as vehicles driving on the roads. Based on the vehicle models provided in the previous section, the state space of all the objects contains states of all the vehicles involved in the decision making problem. For the decision making cycle at time $k$, the state space is:

$$S_k = \{s_k^e, s_k^1, ..., s_k^m\} \tag{14}$$

The subscripts are the time stamp of current decision making cycle. The ego-vehicle's state is represented with superscript $e$ and superscript $1,...,m$ are for the $m$ other vehicles that are not controllable with the decision making algorithm.

$s^e = [x, y, \theta, v, w]^T$. $x, y, \theta$, as also discussed in the vehicle model, represents the longitudinal position, lateral position and yaw angle of the vehicle. These three values provide vehicle pose information. $v, w$ are the linear velocity and yaw rate for describing the motion of the ego vehicle. For the other vehicles,

$$s^i = [x, y, \theta, v]^T,\ i = 1,...,m \tag{15}$$

where $x, y, \theta$ are the vehicle pose information and $v$ is linear velocity of the observed vehicle *i*. The difference in the state tuple between the ego vehicle and other vehicles reveals the partial observability in vehicle decision making problems. Using the vehicle on-board sensors such as IMU, the ego vehicle can sense its yaw rate during the driving. However, due to the limitation of sensors, the ego-vehicle will not be access to other vehicles' yaw rate values. Hence, it generates the problem of predicting other vehicles future trajectory for planning and decision makings while driving in various traffic scenarios. The transition models are the vehicle models provided in the previous section. For the ego vehicle, the transition model is shown in Equations (10) - (11) and for other vehicles, the model is shown in Equations (12) – (13). In this paper, we do not consider the influence of the stochastic process lies in the state transition, hence, we assume the state transitions are deterministic with a probability of transition:

$$T : \mathrm{P}(s' \mid s, a) = 1 \tag{16}$$

The observations states of the agent vehicle and other vehicles are the same with their state tuples with an observation probability of 1. As is to say, the uncertainty of perceptions and sensing is not discussed here, and is out of the scope of this work.

$$Z : \mathrm{P}(o \mid s, a) = 1 \quad (17)$$

Since the OOPOMDP method is employed for the decision making problem with multi-vehicle roundabout intersection traffic scenario, we also formalize the state space and observation space into the OOPOMDP fashion:

- $C = \{c_r, c_v\}$ is the class of objects in the decision making problem. In this problem, we consider two classes of objects, the ego vehicle class $c_r$ for the ego vehicle and other vehicle class $c_v$ on other vehicles in the environment.
- $Att(C)$ are the attributes of objects in different classes. The attribute of the ego vehicle class includes position information $(X, Y, \theta)$ and motion information $(v, w)$. The attribute of other vehicle class includes position information $(X, Y, \theta)$ and velocity $v$.
- $Dom(a)$ is the domain of attributes. The domain of attributes in this work is mainly on the poses of vehicles. For both the ego vehicle class and the other vehicle class, the positional domain is within the area of the roundabout intersection, the heading angle is within the range of $[0, 2\pi)$.
- $A$ is action set. The action is a combination of two finite set of actions, $A = \{a_1, a_2, ..., a_m\} \times \{w_1, w_2, ..., w_k\}$ for linear acceleration and angular velocity. Each action is selected from the set and result in a pair $(a_i, w_j),\ i = 1...m,\ j = 1...k$.

## 8 The penalty Function

The decision making algorithm for round intersection is based on Partially Observable Markov Decision Making Process (POMDP) introduced in previous section. It can also be viewed as an optimal problem that optimizes the expected summary reward in the future horizon. Hence a real-value reward function or penalty function is required for awarding or penalizing the current action. In this work, a penalty style reward function is implemented. Bad action will be penalized which makes it a negative reward, so that the goal of this decision making algorithm is to minimize the overall negative reward (penalization) of the agent traveling along the designated path through the roundabout intersection as shown in Equation (18).

$$a^* = \arg\max_{a \in A} E[\sum_{t=0}^{\infty} \gamma^t R(s_t, a_t) \mid b_0] \tag{18}$$

The goal of this decision making algorithm is to make sure the agent can drive through the roundabout intersection safely, efficiently and comfortably. The requirements for these three perspectives lead to the penalization of the risk of collision, the speed, the acceleration and the well-use of the road. The overall reward function for each step is,

$$\begin{aligned} R(s_t, a_t) = c_1 R_{collision}(s_t, a_t) + c_2 R_{gap}(s_t, a_t) + c_3 R_{velocity}(s_t, a_t) \\ + c_4 R_{target}(s_t, a_t) + c_5 R_{acc}(s_t, a_t) \end{aligned} \tag{19}$$

As shown in (19), it makes up from five different components. Each component is related to partial of requirements for safety, efficiency and comfort.

For the collision rewards, it tells the ego vehicle should not collide with other road users. As a major purpose of deploying autonomous vehicles and self-driving technique, avoiding collision and reducing traffic accidents is the top priority in designing any kind of decision making algorithms. This reward is related to the nearest object, in our work, the nearest vehicle. Every time an action makes the ego vehicle get as close as some threshold safe distance, a large penalty will be given as shown in Equation (20),

$$R_{collision}(s_t, a_t) = \begin{cases} -1000 & \text{if } d(v_e, v) <= d_{safe} \\ 0 & \text{otherwise} \end{cases} \tag{20}$$

Here in the above equation, the notation $d$ refers to the distance between the two vehicles and is given by

$$d(v_e, v) = \sqrt{|x_e - x|^2 + |y_e - y|^2} \tag{21}$$

provides the distance between the center of mass of the agent vehicle (ego vehicle) and that of the nearest vehicle of detection. $d_{safe}$ is determined by the dimension of the vehicles and the speed limit of the road. As illustrated in Figure 5 below, for determining the safety bound of vehicles, a circular range around each vehicle is considered, in which case the boundary will not be influenced by the heading direction of vehicles. Therefore, the safety threshold distance is a summary of the threshold distance as well as two radiuses of the circular boundaries generated from the vehicles' dimensions, as shown below.

$$d_{safe} = d_{threshold} + R_{e,boundary} + R_{boundary} \tag{22a}$$

$$d_{threshold} = \frac{v_{\lim}^2}{2a_{\max}} \tag{22b}$$

$$R_{e,boundary} = \sqrt{W_e^2 + (L_e / 2)^2}, R_{boundary} = \sqrt{W^2 + (L/2)^2} \tag{22c}$$

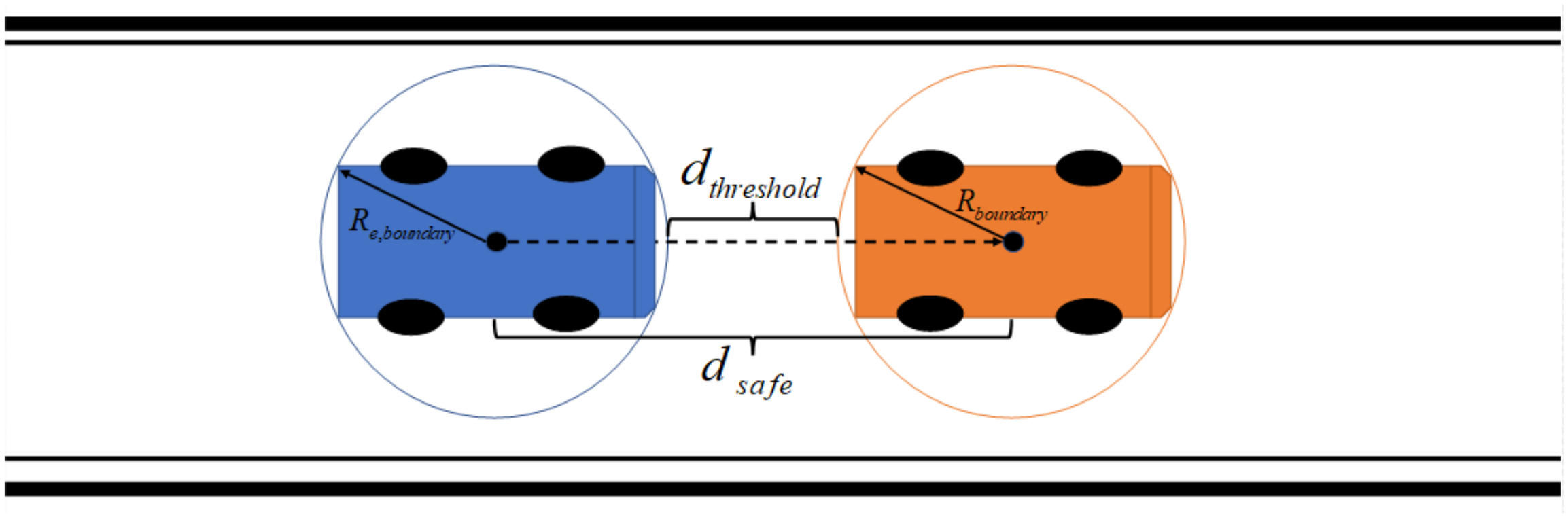

Figure 5 Diagram of Reward (Cost) for Collision

In Equations (22), $v_{\lim}$ is the speed limit on a given road segment, and $a_{max}$ is the maximum allowed deceleration of a vehicle for which the passengers in the vehicle will not feel uncomfortable and $d_{threshold}$ shows the emergency distance when the latter vehicle (the blue vehicle in the back) needs to make an emergency brake to fully stop in order to avoid collision with the front vehicle (the orange vehicle). $W$ represents the width of the vehicles, $L$ is the length of the vehicles.

As is discussed in [21], the ego vehicle is penalized for its speed being lower than the speed limit of the current road segment as the vehicle is always expected to drive at the limit speed of the current road segment for efficiency. It is also penalized if it exceeds the speed limit. The penalty for exceeding the speed limit is larger than being lower than the limit speed. Also, in this part the penalty for not discomfort is also introduced.

$$R_{velocity} = \begin{cases} C_{exceed} \dfrac{|v_e - v_{des}|}{v_{des}} & \text{if } v_e > v_{des} \\ C_{lower} \dfrac{|v_e - v_{des}|}{v_{des}} & \text{if } v_e < v_{des} \end{cases} \tag{23}$$

$C_{exceed}$ and $C_{lower}$ are negative constant coefficients at the situation where ego velocity $v_e$ is larger than or smaller the desired velocity $v_{des}$ respectively. The desired velocity is determined according to the largest tolerance lateral acceleration for passengers in the vehicle if the vehicle is turning and follows the road speed limitation if the vehicle is traveling on straight road as,

$$v_{des} = \begin{cases} \sqrt{a_{y,\max} / \kappa} & \text{if turning} \\ v_{limit} & \text{otherwise} \end{cases} \tag{24}$$

The velocity reward will drive the ego vehicle towards desired velocity if there is no risk of collision with other vehicles. Once, there is a chance of collision, the velocity reward will be dominated by the penalty caused by collision risk due to the difference of scales of these two rewards.

As for the passengers in the passenger vehicle, the maximum tolerance of the acceleration is

$$a = \sqrt{a_x^{\ 2} + a_y^{\ 2}} <= a_{\max} \tag{25}$$

Therefore, once the overall acceleration is larger than $a_{\max}$, a penalty for discomfort is introduced in the reward function for causing discomfort to the passengers in the vehicle. The longitudinal acceleration part $a_x$ is the currently selected action from the action set. The lateral acceleration $a_y$ is derived from the lateral motion of the vehicle as

$$a_y = v^2 / r = v^2 \kappa \tag{26}$$

where r is the radius of vehicle turning and $\kappa$ is the corresponding curvature.

$$R_{acc} = \begin{cases} -100 & \text{if } a >= a_{max} \\ 0 & \text{otherwise} \end{cases} \tag{27}$$

Since the vehicle is following a designated path, the target area is also pre-determined on the path. Once the vehicle is within a distance of the target point of the path, it will be rewarded.

Apart from the commonly seen reward settings as can be seen in other works, a gap reward is also proposed for the driving efficiency on road in this decision making algorithm. Intersections are normally the most crowded traffic scenarios in urban traffic, especially round intersection. In round intersections, vehicles are merging together from different road segments. Besides the requirement of safety, efficient use of the space on road is also necessary in order to reduce the chance of traffic jam. The gap reward aims to drive the agent vehicles to follow the proceeding vehicle at a certain distance. A commonly accepted 3-second rule for vehicle following is adopted for generating the desired gap between the agent vehicle with the preceding vehicle as

$$v_{3s} = \max\left\{v_e - 3 * b_{\max}, 0\right\} \tag{28a}$$

$$d_{desired} = \frac{v_e^2 - v_{3s}^2}{2a_{\max}} \tag{28b}$$

$$R_{gap} = c_{gap} \mid d(v_e, v) - d_{desired} \mid \tag{28c}$$

Here in the Equation (28a), $v_{3s}$ is the target velocity when acting emergency full brake at the maximum brake acceleration $b_{\max}$ and the $d_{desired}$ is the preferred gap between the ego vehicle and the nearest preceding vehicle. A reward will be gained if the distance is different from the desired gap between vehicles.

## 9 Improving Decision Making with Policy Prediction

Decision making and planning for autonomous vehicle in urban traffic scenario has always been a challenging problem due to the complexity of urban traffic. Another challenge is the tracking of vehicle driving motion in the urban area that largely influences the performance of the decision making algorithm. There are several research works that are conducted to improve the vehicle planning algorithm performance: [22] proposes the method of adjusting the vehicle model parameters (IDM model parameters) to emulate different style of driving, and tries to improve the decision making the algorithm for highway autonomous driving performance; [23] proposed a game-theoretic based planning algorithm for roundabouts and traffic circle that applies the game-theoretic based planning framework on all the involved vehicles in the traffic scenario. [24] proposed a method that uses hidden mode stochastic hybrid system to model different human driving behaviors to assist the performance of decision making in driver assistance system. Yet, there are rare works for improving the autonomous vehicle decision making problem for round intersections with multiple vehicles involved. Hence, in this section, a policy prediction method based improvement on autonomous vehicle decision making is proposed and is one of the main contributions of this chapter.

The planning and decision making problem can be regarded as a receding horizon optimization problem starting from current time and extending to a future time based on the horizon of the optimization problem. The ego vehicle obtains an observation on all the other surrounding vehicles at time t, noted as $o_t$ and tries to find an optimal action such that the total reward is maximized as

$$a^* = \arg\max_{a \in A} E[\sum_{t=t}^{t+k} \gamma^t R(s_t, a_t)] \tag{29}$$

We do not have full knowledge of all the future states along the planning horizon. It is essential to have a good prediction over the future state trajectories and the state transition model needs to have good precision on tracking the future potential trajectories.

For decision making and planning of autonomous vehicles, the vehicle trajectory tracking over a future horizon can be done with the state transition model as described in Equations (12) – (13).

For highways, it is rather simple since the vehicles tend to not have rough behaviors during driving, for example, accomplishing multiple lane-changing behavior within a short time period. The model in Equation (12) can provide a good trajectory tracking for the decision making problem on a highway. The case is different for planning in the urban traffic environment for intersections, especially round intersections. The vehicle needs to change the direction of driving in a short period of time due to the geometric features of intersections and round intersections when crossing an intersection or merging into a round intersection. The current observations of other vehicles does not provide enough information on how the vehicle will be moving in the future time horizon for planning.

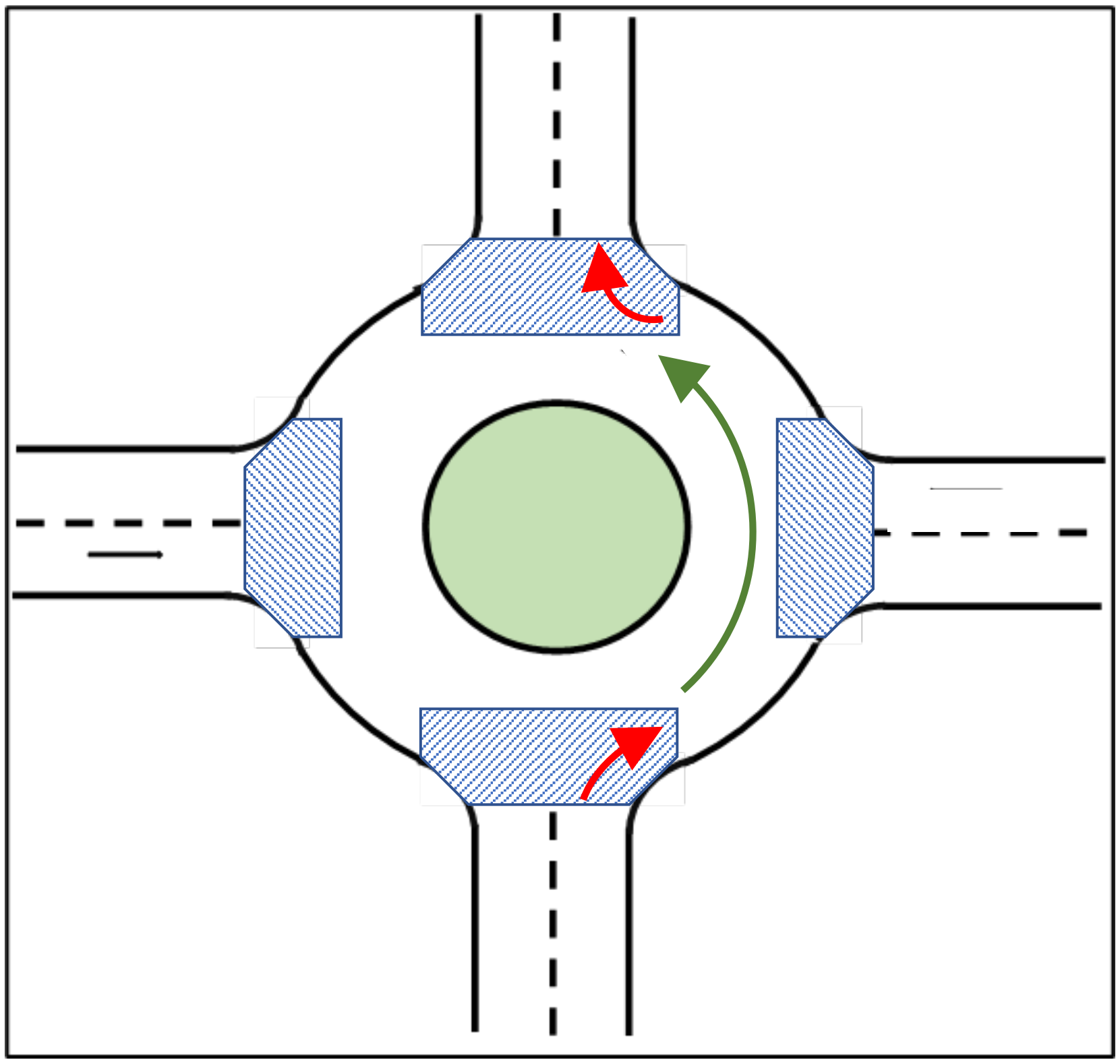

Figure 6 Diagram of Vehicle Passing a Round Intersection, Shadow Area Refers to the Area where Vehicles Enter/Exit the Round Intersection

As a demonstration example, the path of a vehicle passing a round intersection is shown in Figure 6. The shadowed areas in the figure are the areas in which a vehicle is entering or exiting the round intersection. The vehicle first travels through a shadowed area to enter the round intersection, takes a turn in clock-wise direction. After passing the shadowed area to enter in the round intersection, the vehicle follows the lane area and takes a constant yaw angle turning in the counter-clockwise direction. Reaching the target connection point to exit, the vehicle now, takes a clock-wise turn to exit the round intersection and drive on the straight road again. Through the whole procedure of passing the roundabout intersection, the vehicle shows a discontinuous motion in driving direction and the yaw rate keeps changing along the path and makes it hard for the vehicle motion model (12) to track given a single initial condition.

In a decision making cycle, the planning starts from some initial state sampled from the belief state space at current time, noted as $t_0$, and carries out a forward simulation to generated potential trajectories of the states for optimizing the expected cumulative reward along the optimization horizon $t_0 \sim t_0 + k$, with $k$ being the decision making horizon. Directly implementing the vehicle motion models in Equation (12), even with the knowledge of yaw rate at $t_0$, the true trajectory deviates a lot from the simulated trajectory that tracks the potential vehicle trajectory on the optimization horizon. The model alone is not enough to describe the vehicle state transitions under the roundabout intersection traffic scenarios, as shown in Figure 7. In Figure 7, the true trajectories are not successfully tracked by directly implementing the state transition Equations from the initial state. The forward simulation path is calculated for 100 steps and the arrows show the direction of vehicle moving or the direction of simulation. This will cause a lot of issues when making decisions for the ego vehicle to pass the round intersection safely and efficiently. This issue will also affect the solution of the POMDP problem based on Monte-Carlo Tree search since Monte- Carlo simulation for the state particles is the core procedure for the tree search algorithm.

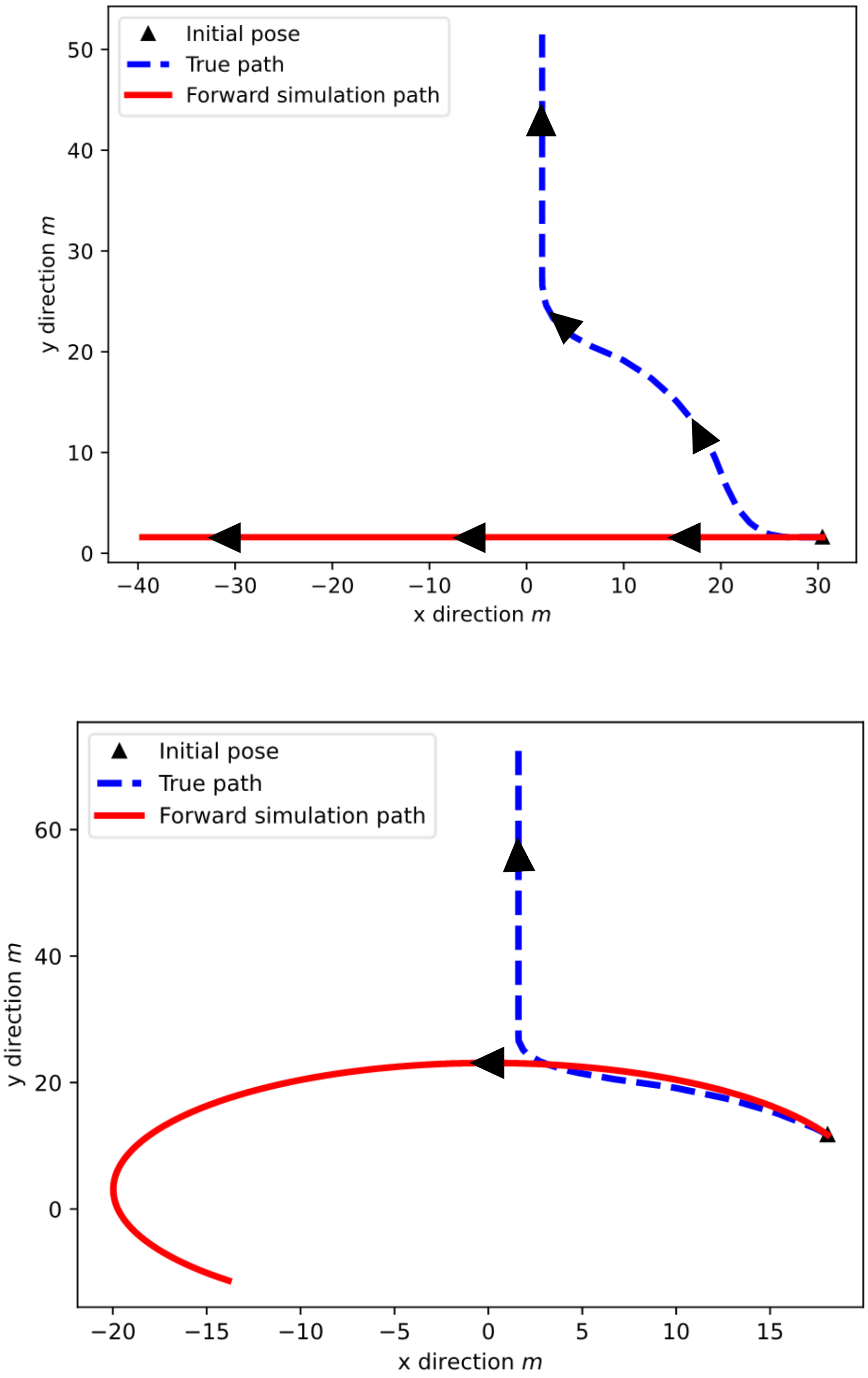

Figure 7 Forward Simulated Path Obtained by Using the State Transition Model Directly

To overcome the issue caused by discontinuity of vehicle motion a policy prediction based decision making method is proposed here such that the state transition of the vehicle can be better tracked and simulated during the decision making cycle. The policy prediction method is based on change point policy prediction method. This paper mainly focuses on the traffic scenario of roundabout traffic circle and an abstract policy set is applied here for the decision making algorithm. This method utilizes the recorded history of the decision making algorithm based on OOPOMDP so that it avoids additional memory cost.

**10 Augmented Objective State and Policy Based State Transition**

In this section, we present an augmented objective state based on the ego vehicle's observations of other vehicles so that it can be used for the policy based state transition. One of the features of the OOPOMDP is that the state space and the observation space are factored into an object space, so that the state space, including states, observations and belief states are all factored accordingly on the objects from the objects set. Also, the belief update is per object.

The original setting of the observation state is the same with the state of other vehicles, that is

$$s_t^i = o_t^i = [x_i, y_i, \theta_i, v_i]^T \tag{30}$$

where, the observation state $o_t^i$ is the real observation made by the agent from the environment. It doesn't contain any of the policy information so that the states sampled from the updated belief states will directly accomplish state transition using the model (12) during the decision cycle. This will lead to the issue of simulated path deviates from the true path and cause the inaccuracy of the decision making. Here, an attribute is added to other vehicle class $c_v$ for the most recent policy $\pi_i$, and augments the state space and observation space into

$$s_{t,a}^i = o_{t,a}^i = [x_i, y_i, \theta_i, v_i, \pi_i]^T \tag{31}$$

where $\pi_i$ is the most likely policy that the observed vehicle is executing at the time that current observation is made for vehicle *i*. Denote the policy prediction method as

$$F : h \rightarrow \pi_t^i, i \in \{1, 2, ..., n\} \tag{32}$$

Here $h$ is the current history in the POMDP problem, and as shown in previous sections, is a series of actions and observations made by the agent,

$$h_t = \{b_0, a_0, o_1, a_1, o_2, ..., a_{t-1}, o_t\} \tag{33}$$

According to the feature of OOPOMDP, the observations of other vehicles can also be indexed by object in the history. So, we can use it for policy prediction method for each of the vehicle involved in the traffic scenario. $\pi_t^i \in \Pi$ is a policy for vehicle *i* selected from the pre-determined policy set. For every decision cycle, we first obtain the history of observations on vehicle $i$ and implement the policy prediction method to find the policy over the last segment of the vehicle, then augment the observations state and state of vehicle *i* with the potential policy so that the policy can be used for state transition during the Monte-Carlo Tree search. The state transition equation of other vehicles is now modeled as

$$S_{t+1} = T_v(S_t, \pi_t) \tag{34}$$

which is

$$x(t+1) = x(t) + v(t)\cos\theta(t)\Delta t \tag{35a}$$

$$y(t+1) = y(t) + v(t)\sin\theta(t)\Delta t \tag{35b}$$

$$v(t+1) = v(t) \tag{35c}$$

$$\theta(t+1) = \theta(t) + w(\pi_i, \hat{\theta}_{\pi_i})dt \tag{35d}$$

The vehicle model (12) now becomes a constant velocity and turn rate model for vehicle trajectory tracking. The yaw rate is generated based on current policy $\pi_i$ and parameter $\hat{\theta}_{\pi_i}$ that will help with the vehicle trajectory tracking in the decision making cycle and improve the performance of the decision making based on OOPOMCP algorithm.

Since the decision making algorithm is mainly used for vehicles passing the roundabout intersection, we take advantages of its geometry to simplify the generation procedure of vehicle yaw rate based on the policy. According to the geometric features of the roundabout intersection, we know that a round intersection has mainly three types of areas: straight road segments, enter/exit area and round area. The vehicle yaw rate will also be determined by the road curvatures in these three different types of areas, with the equation:

$$w = v / r = v \cdot \kappa \tag{36}$$

where $r$ is the turning radius of the vehicle trajectory and $\kappa$ is the road curvature.

To illustrate the difference in utilizing this method, a forward simulation trajectory is generated by the policy based state transition model and shown Figure 8. From Figure 8, we can see that the policy based simulated path is very much similar to the true path of a vehicle passing the round intersection where only given the initial state is given. In the simulated path generation, we assume that the policy of entering the round intersection is given such that the vehicle yaw rate is generated based on the policy, so the vehicle is turning in a clockwise direction to enter the round intersection instead of keeping the original heading angles and yaw rate from the straight lane. In this case, the decision made by the agent vehicle will be closer to the real situation and improves the performance of the decision making algorithm. The methods for decision making based on policy prediction state update is shown in Table 3.

| Algorithm: Decision Making with Policy Prediction |
| --- |
| **Procedure** Search $(h)$ :<br>$b \leftarrow$ Prior<br>$s \leftarrow$ InitialState<br>$\pi_0 \leftarrow$ DefaultPolicy<br>Repeat:<br>$\hat{s} \sim$ SAMPLE$(b)$<br>SIMULATE$(\hat{s},\{\},0)$<br>Until Timeout()<br>$a \leftarrow \arg\max_a V(ha)$<br>if $length(h) > L$:<br>for $obj_i \in obj$:<br>$\pi_i, \hat{\theta}_{\pi_i} = F(h)$<br>else:<br>$\pi_i = \pi_0$<br>$s', o, r \leftarrow \varepsilon(s,a)$<br>$o_a =$ Augment(o)<br>$b' \leftarrow$ UPDATE$(b,a,o_a)$<br>**end Procedure** |

Table 3 Method for Decision Making with Policy Prediction

In this method, the SIMULATE, SAMPLE and UPDATE methods are all the same from the OOPOMCP algorithm shown in Table 2 and used for solving the OOPOMDP problem.

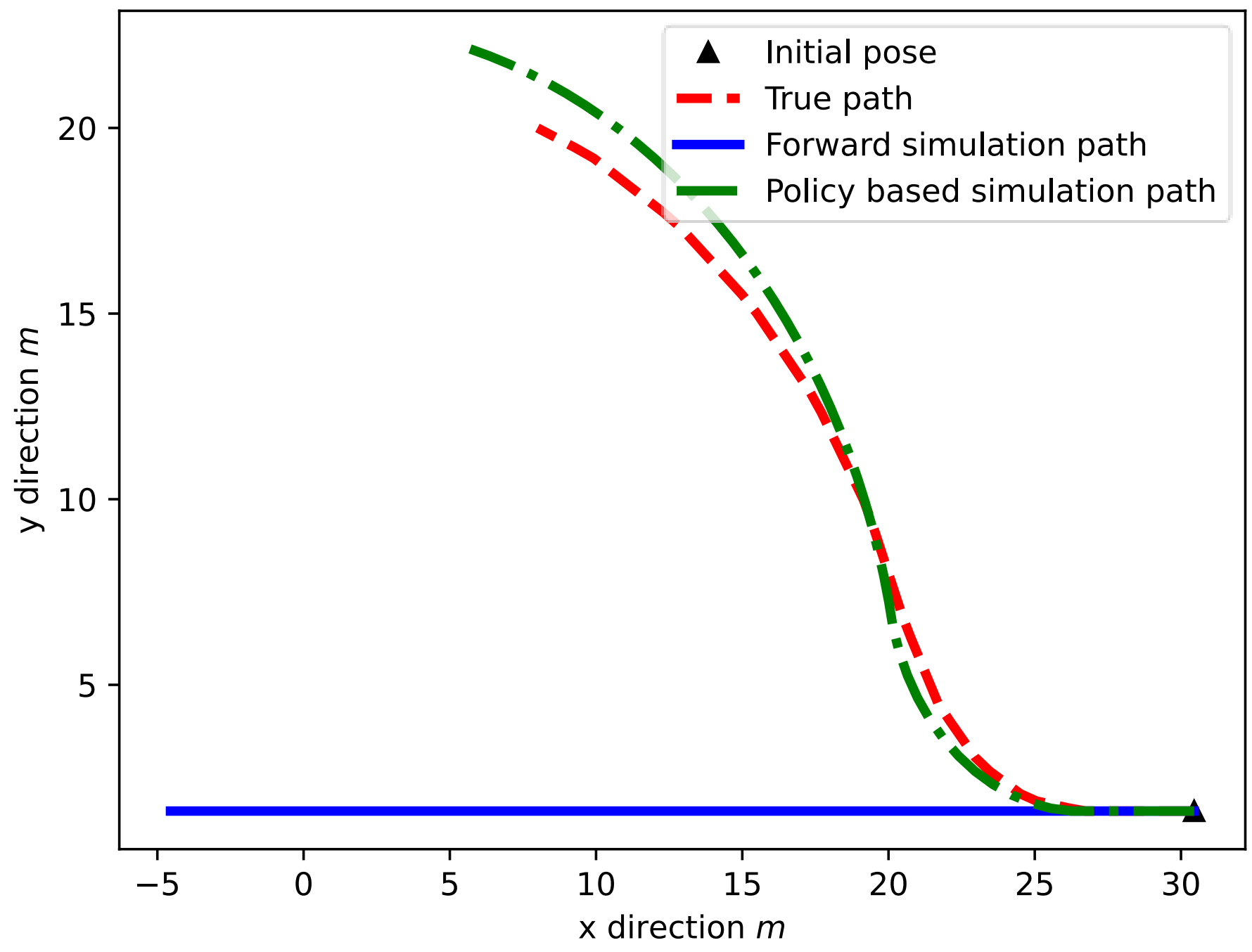

Figure 8 Policy Based Forward Simulation

Besides using the policy based state transition in the decision making problem of vehicles in round intersection, the vehicle control of the autonomous vehicle which is about the vehicle jerk for velocity control is also considered. Since the reward function can measure how good it is for current selection of an action from the action set, it is able to measure the rewards caused by velocity that exceeds the speed limit or exceed the maximum lateral acceleration that causes the discomfort. However, it is not capable of measuring the change of acceleration since the reward function does not have the access to previous actions. Therefore, the previous linear acceleration is stored for comparison with current acceleration determined by the tree search. We approximate the jerk of the vehicle to be:

$$J_t = |\Delta a| = |a_t - a_{t-1}| \tag{37}$$

Based on this jerk indication $J_t$, an acceleration re-selection is made to make sure that the acceleration change does not exceed a maximum allowed jerk so that the vehicle speed change is smooth and passengers will feel comfortable.

$$a_t = \begin{cases} a_t & \text{if } J_t <= J_{\max} \\ a_{t-1} + J_{max} & \text{if } a_t - a_{t-1} >= J_{\max} \\ a_{t-1} - J_{max} & \text{if } a_t - a_{t-1} <= -J_{\max} \end{cases} \tag{38}$$

In this way, it is certain that the jerk of the autonomous vehicle will not be too large.

## 11 Simulations and Discussion

The simulation of decision making for autonomous vehicles passing a round intersection is done for test and validate the decision making algorithm with policy based state transition. The traffic scenario in the simulation is a typical 4-way round intersection. Vehicles enter the scenario from the straight lane segment and move towards its destination area. The diagram of the simulation environment is shown in Figure 9.

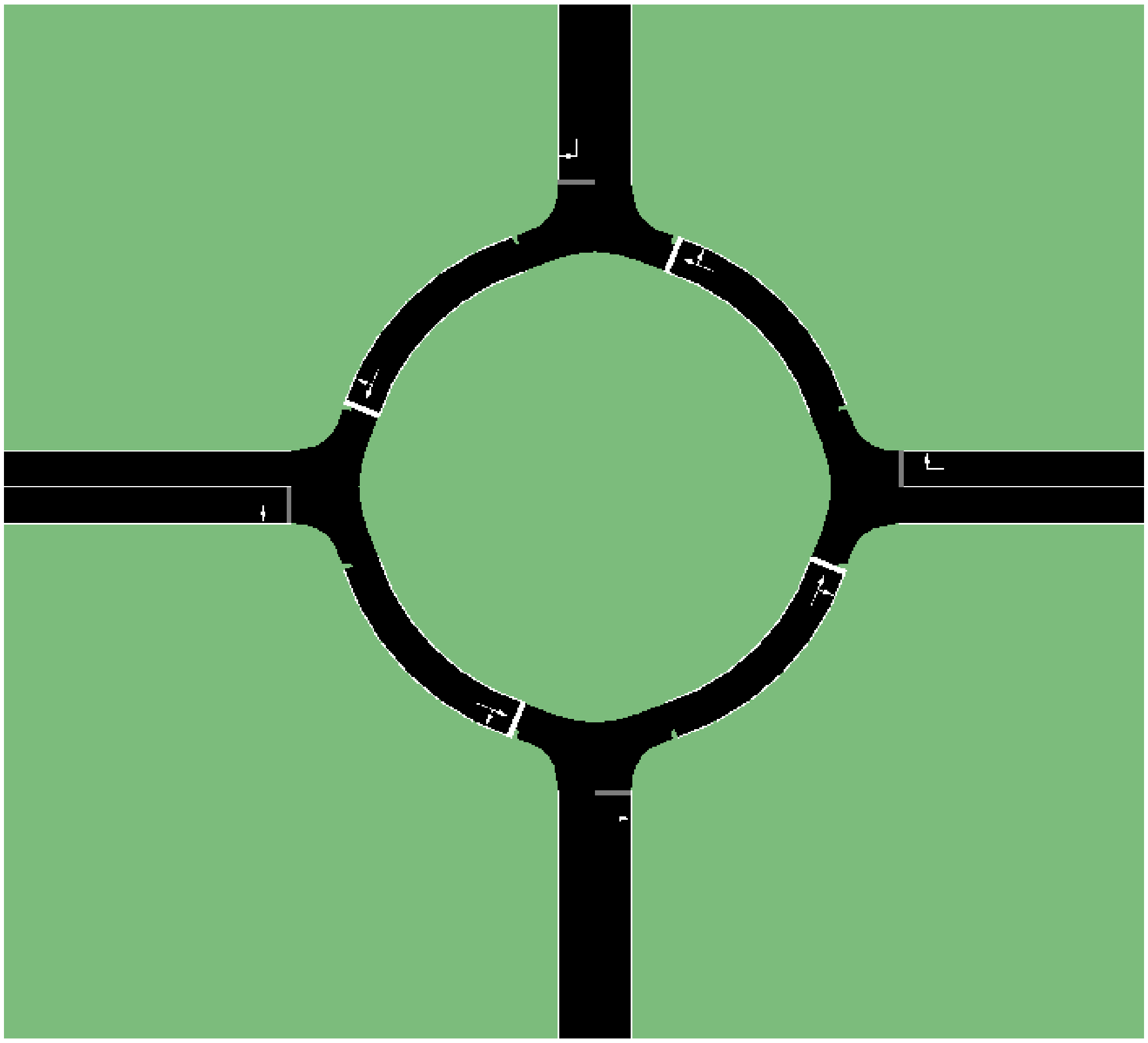

Figure 9 SUMO Simulation Environment

For simplifying the test environment, we assume the roundabout intersection is a circle so that the road curvature within the round intersection is a constant: $\kappa = 1/r$ where $r$ is the radius of curvature of the lanes in the round intersection. For the simulation, vehicles will drive on the right side of the road on straight lane segment and in the counter-clockwise direction of within the round intersection.

For the OOPOMCP algorithm to solve the Objective-Oriented POMDP Problem, the parameters for the tree search are:

- Planning time for each step: 1.0 s

- Maximum search depth in the tree: 100
- Exploration constant in the UCB1 value: 2.0
- Discount factor of the cumulative expected reward: 0.99

In the action set, a discrete linear velocity is given by

$$Acc = \{-3.0,\ -2.0,\ -1.0,\ 0,\ 0.5,\ 1.5,\ 2.5\}\ m/s^2$$

The yaw rate of the vehicle can be determined according to the road curvature with the linear velocity at different road segment for a round intersection. As for a single lane round intersection with straight road segment connecting to it, the yaw rate appears to have 3 features:

- In the straight lane segments, the vehicle's yaw rate keeps at zero as it keeps driving straight.
- In the round intersection part, the vehicle will drive along the road at the yaw rate of $w = \frac{v}{R}$ where $R$ is the radius of the round intersection and $v$ is the vehicle linear velocity, and this is a positive value when setting the counter-clockwise direction is positive direction.
- When the vehicle is entering/exiting the round intersection, the yaw rate $w'$, is some negative value when the counter-clockwise direction is taken as the positive direction. In the simulation, since the entering/exiting process only takes a very short time of period, the yaw angle change is approximated with a yaw rate determined by the process of uncontrolled simulation vehicles entering the round intersection.

In the simulation, the round intersection has a radius of curvature of 20 *m,* hence the road curvature in the roundabout intersection is 1/20. Additionally, based on the test data, the curvature of the entering/exiting area in the simulation is approximated to be 0.15 and will be used in the transition model. Apart from those parameters in the action set, parameters for the reward function in the decision making problem are listed below. Currently, all the factors in the total reward is equally weighted, hence the coefficients:

$$c_1 = c_2 = c_3 = c_4 = c_5 = 1$$

The collision reward depends on the vehicles dimension. In this test, the vehicles have same dimension, which are $L = 5.0\ m, W = 1.8\ m$, which is the default settings of the passenger vehicles in SUMO simulation.

The overall maximum acceleration required in Equation (16) is set to be 4.0 $m/s^2$, and the maximum lateral acceleration caused by vehicle turning is set to be 2.0 $m/s^2$. The cost due to exceed the desired velocity in Equation (12) is $C_{exceed} = -100$ for penalize the unsafe maneuvering of exceeding the maximum allowed velocity, and the cost for being lower than the desired velocity is $C_{lower} = -10$. The maximum brake acceleration is the lowest acceleration from the action set that is -3.0 $m/s^2$ and $c_{gap} = -10$ .

The simulation results compares the performance of the decision making algorithm with decision making using OOPOMDP without any policy based state transition and the SUMO simulation system built in vehicle model, which is an Intelligent Driver Model (IDM) that execute lane following. The results compare the total travel time, total reward and if there is emergency brake acted by other vehicles in the system.

First a two-vehicle scenario simulation is done and result are presented below. The diagram of two vehicle scenario is presented in Figure 10. The autonomous vehicle colored with green is traversing the roundabout intersection from the bottom of the intersection and trying to reach the destination points up in the figure. The simulation results for this two vehicle scenario are summarized in Table 4.

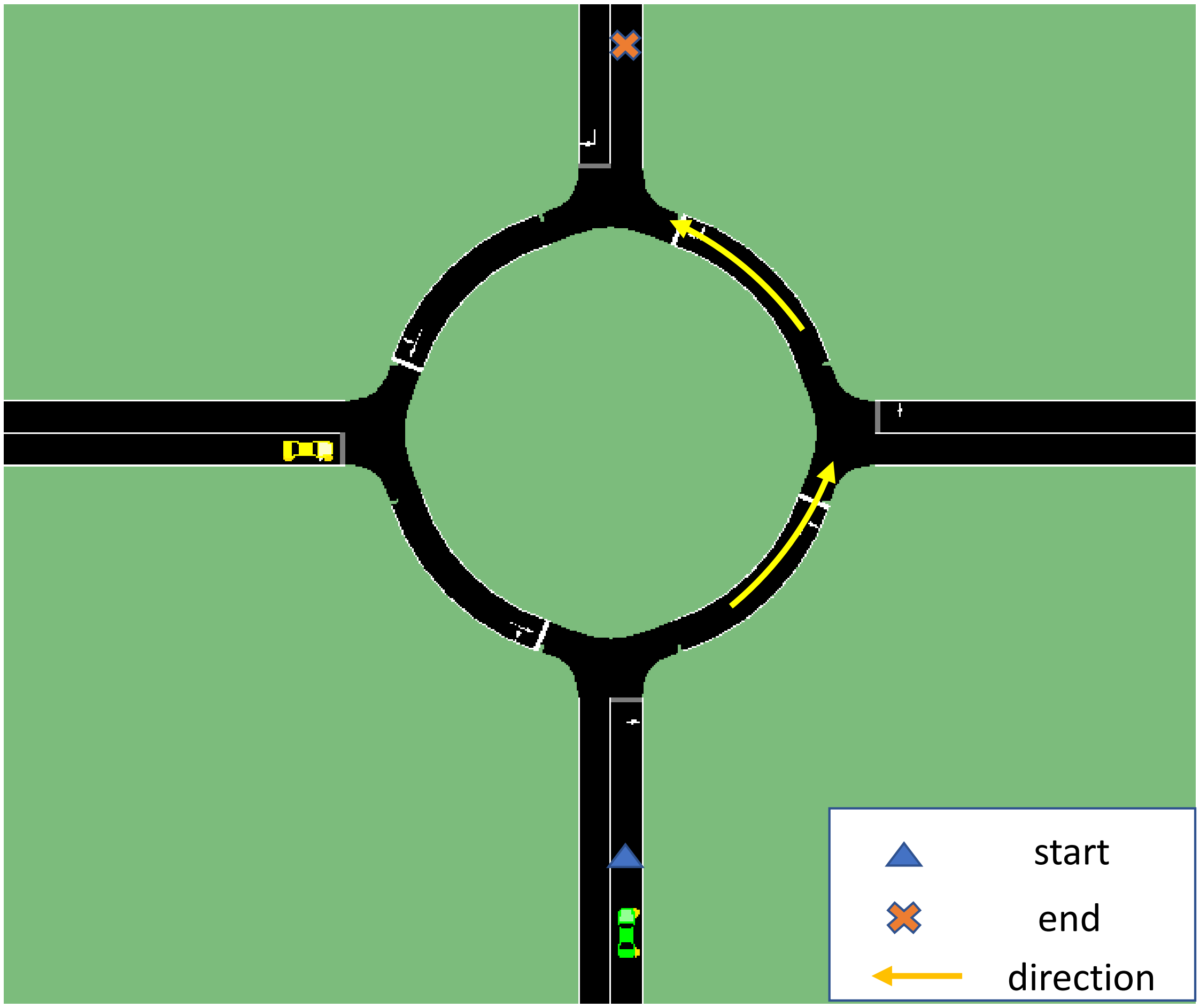

Figure 10 Two-vehicle Scenario of Passing the Intersection

| | Total Reward | Total Travel time | Emergency Brake |
|---|---|---|---|
| OOPOMDP with policy based state transition | -527.9 | 21s | No emergency brake performed |
| OOPOMDP based decision making | -593.674 | 23s | Emergency brake performed by another vehicle |
| System driver | | 23s | No emergency brake performed |

Table 4 Simulation Result of the 2-vehicle Scenario

From the result above, we can find that the OOPOMDP with Policy based state transition method proposed in this work achieves higher reward compared with the one which directly implements the state transition model proposed in Equation (12). With the policy based state transition, the ego vehicle can better predict the future trajectory of the surrounding vehicles and control the speed and direction to reach destination in a shorter time. Also, since it avoids the risk of colliding with other vehicles, no emergency brake or aggressive maneuvers need to be made by other vehicles while they are following the lane with IDM model. Also, the shorter time achieved when comparing to the system driver proves the efficiency of implementing such decision making algorithm for the autonomous vehicles.

Another simulation is carried out for a multi-vehicle scenario. Eight vehicles in total are involved in the simulation with different departure times. Hence, they will interact within the region of the round intersection, as shown in Figure 11.

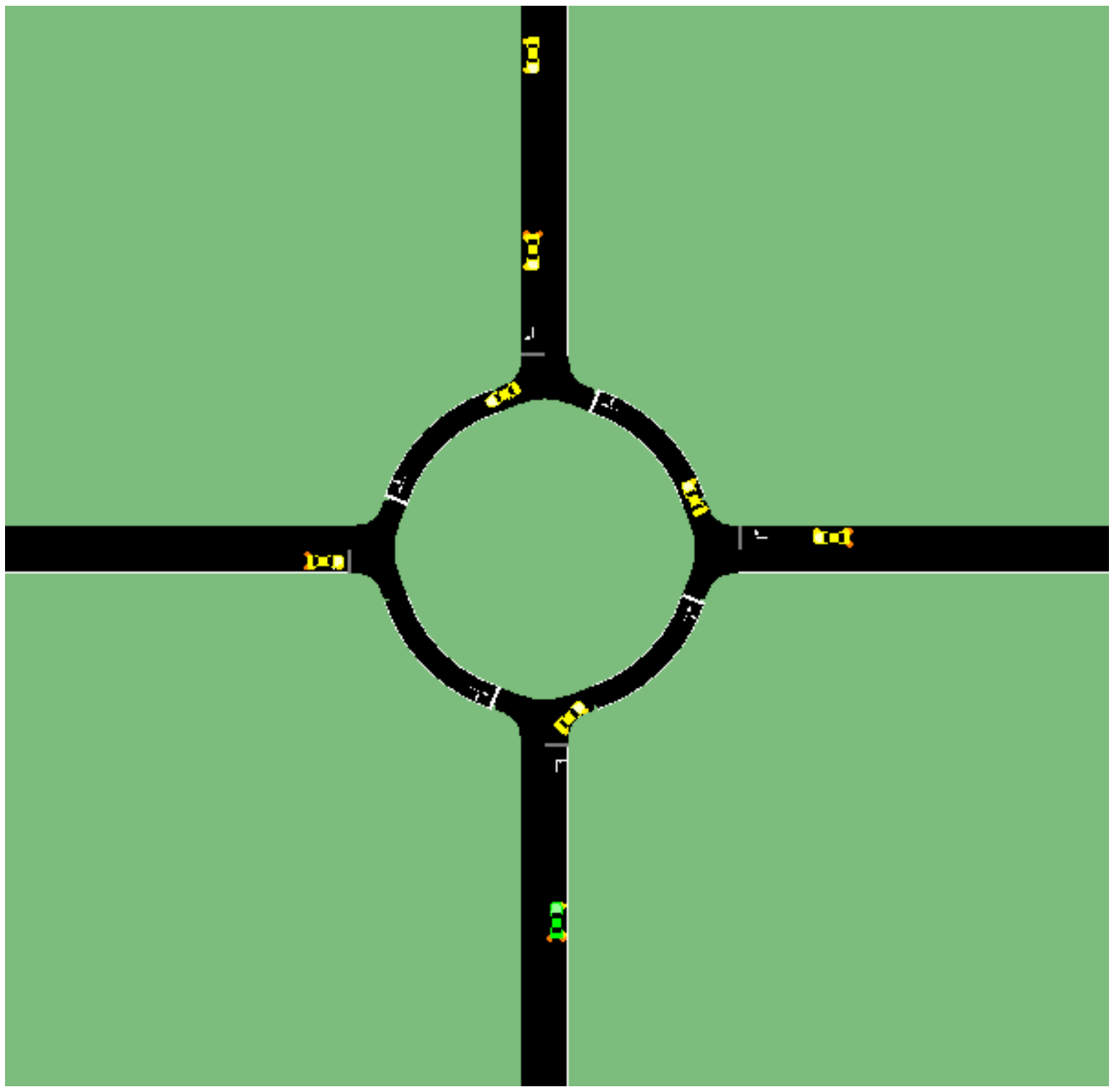

Figure 11 Multi-vehicle Scenario

Still the green vehicle shows the ego vehicle that deploys the decision making algorithm and the test results are summarized in Table 5.

| | Total Reward | Total Travel time | Emergency Brake |
|---|---|---|---|
| OOPOMDP with policy based state transition | -450 | 20.9s | No emergency brake performed |
| OOPOMDP based decision making | -14516.3 | 22s | Potential Collision |
| System driver | | 33.7s | No emergency brake performed |

Table 5 The Test Result of Multi-vehicle Simulation

In this simulation, the decision making with policy based state transition is able to achieve the goal of traversing the round intersection safely and efficiently without any potential collision and takes advantage of road space to drive fast. Yet, we see that the OOPOMDP based decision making has large penalty, that is because a potential collision that generates a very large penalty for the possible collision happens due to the inaccuracy of predicting other vehicles trajectories as shown in Figure 12.

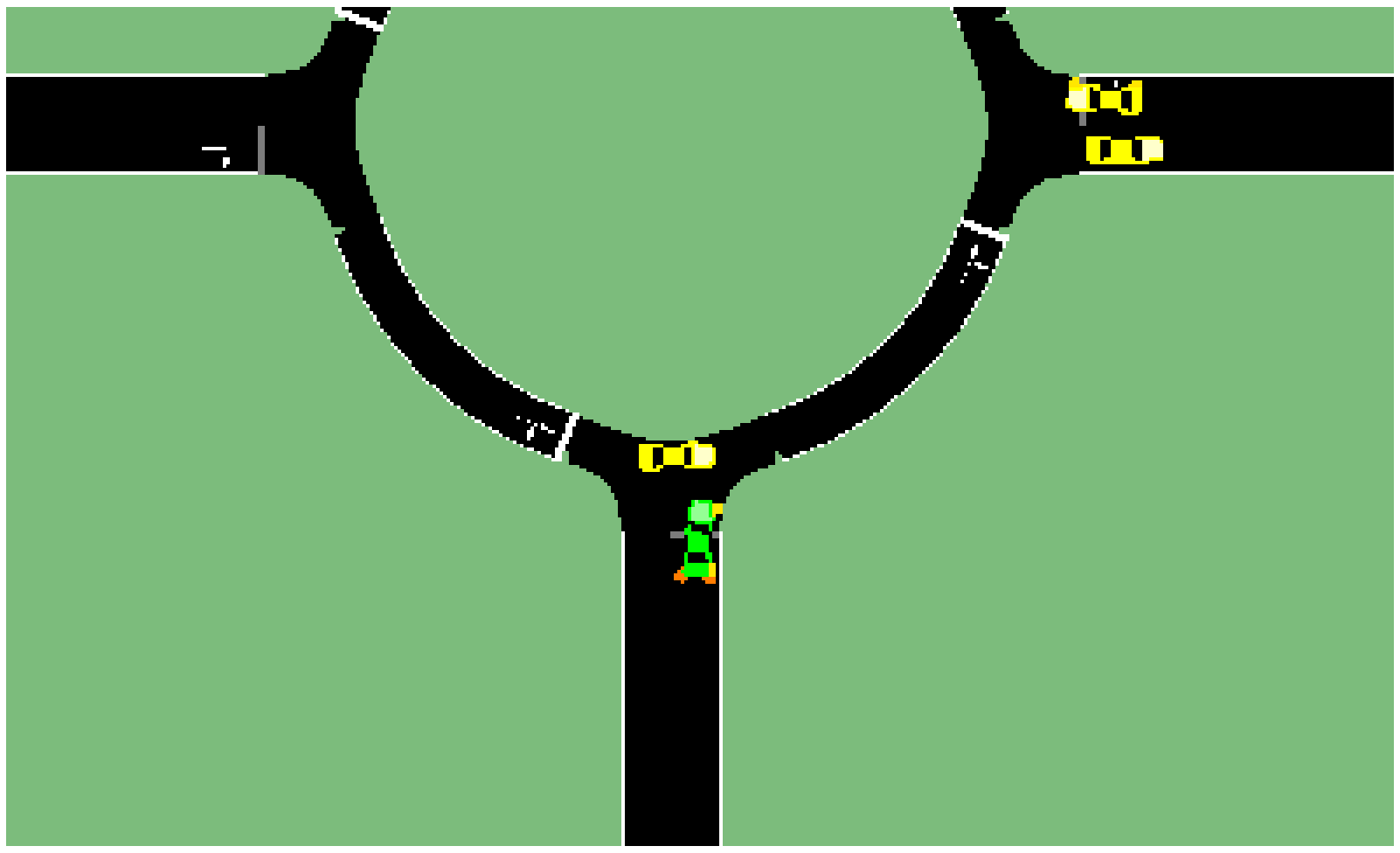

Figure 12. Potential Collision

The long traversal time of the system driver is due to the right of way, the ego vehicle has to wait for other vehicles in the round intersection to pass. Thus, the waiting time is very long compared with the result of implementing the decision making algorithm.

## Conclusions and Recommendations

In this paper, methods for decision making and planning of autonomous vehicles were introduced and discussed along with simulation result for validation of the algorithms. Decision making for autonomous shuttle aiming to solve last mile problem, decision making for autonomous vehicles for round intersection scenario and velocity planning methods were proposed.

However, when the traffic scenarios are complex with multiple road users involved, the rule based decision making algorithm is not robust enough for planning behavior for autonomous vehicles. POMDP is a powerful tool for solving the sequential decision making problem with uncertainties and states that are not fully observable. The OOPOMDP algorithm provides a method for factoring object states and observations individually, and makes the belief update more efficient and low cost. Utilizing the feature of OOPOMDP, a policy based state transition is employed for the decision making algorithm. Since the OOPOMCP stores the history of all the involved agents and environment objects, the agent gets to determine the most recent policy of other vehicles based on the policy prediction method introduced. It largely improves the performance of the decision making for the traffic scenario of round intersections. The settings of reward function is also based on the driving requirements of the autonomous vehicles, to drive safely and efficiently.

Platooning or convoying of vehicles in the form of adaptive and cooperative adaptive cruise control [25], [26] on highways is a topic of high research attention. More recent work focuses on similar cooperative driving in urban roads including cooperative handling of an intersection by a convoy of cooperating vehicles [27]. While there are results for signalized intersections, corresponding results for roundabouts and traffic circles are missing. The approach in this paper can be useful for cooperative handling of roundabouts by a convoy of connected and autonomous vehicles. Active safety control systems like yaw stability controllers [28], [29] are also important as the vehicles in the roundabout follow a circular path and yaw stability problems may occur due to weather conditions or emergency maneuvers.

**Acknowledgments**

The authors thank the support of the Automated Driving Lab at Ohio State University. The authors acknowledge the partial support through the Smart Campus organization of the Ohio State University in support of the Smart Columbus project.